\documentclass[pre,twocolumn,amsmath,showpacs]{revtex4}
\usepackage{graphicx}
\usepackage[dvips]{color}
\usepackage{bm}

\def\DLSF{G}
\def\ADLSF{\hat{G}}
\def\size{N}
\def\lambdamax{\lambda_N}
\def\lambdaneg{\lambda_\infty}
\def\totpartnum{p}
\def\sitepartnum{n}
\def\hop{\eta}
\def\time{t}
\def\position{y_\time}
\def\GF{Z}
\def\gam{\gamma}
\def\parameters{\Omega}
\def\f{\lambda}
\def\fall{\f_\size(\gam;\parameters)}

\def\GFall{\GF_{\time,\size}(\gam;\parameters)}
\def\Fall{\f_\infty(\gam;\parameters)}
\def\a{a_{\parameters}}
\def\b{b_{\parameters}}
\def\partdensity{\rho}
\def\smallT{T}
\def\bigT{\hat{\smallT}}
\def\C{C}
\def\d{d}
\def\eigen{\xi}
\def\c{z}
\def\q{q}
\def\B{B}
\def\aD{a_{\partdensity}}
\def\bD{b_{\partdensity}}
\def\lconstD{c_\partdensity}

\begin{document}

\title{Numerical Method for Accessing the Universal Scaling 
Function for a Multi-Particle Discrete Time Asymmetric Exclusion Process}

\author{Nicholas Chia}
\author{Ralf Bundschuh}
\email{bundschuh@mps.ohio-state.edu}

\affiliation{Department of Physics,
Ohio State University, 191 W. Woodruff Ave.,
Columbus, OH  43210, USA}

\date{\today}

\begin{abstract}
In the universality class of the one dimensional Kardar-Parisi-Zhang 
surface growth, Derrida and Lebowitz conjectured the universality of 
not only the scaling exponents, but of an entire scaling function. 
Since Derrida and Lebowitz's original publication~[PRL \textbf{80} 
209 (1998)] this universality has been verified for a variety of 
continuous time, periodic boundary systems in the KPZ universality 
class. Here, we present a numerical method for directly examining the 
entire particle flux of the asymmetric exclusion process (ASEP), thus 
providing an alternative to more difficult cumulant ratios studies.
Using this method, we find that the Derrida-Lebowitz scaling function
(DLSF) properly characterizes the large system size limit ($N\to\infty$) 
of a single particle discrete time system, even in the case of very 
small system sizes ($N\leq 22$). This fact allows us to not only verify 
that the DLSF properly characterizes multiple particle discrete-time 
asymmetric exclusion processes, but also provides a way to numerically 
solve for quantities of interest, such as the particle hopping flux. 
This method can thus serve to further increase the ease and 
accessibility of studies involving even more challenging dynamics, such 
as the open boundary ASEP. 
\end{abstract}


\maketitle

\section{Introduction}
The Kardar-Parisi-Zhang (KPZ) equation~\cite{KPZ} describes a rich 
variety of processes such as surface growth~\cite{kand90,krug91a}, 
directed polymers~\cite{brun00,bund96,lass97}, and 
avalanches~\cite{povo03,chen02}. 
And accordingly, the massive array of literature (see 
\cite{Haplin-Healy,Meakin,Krug} and references therein), on the topic 
of KPZ theory reflects the central role the KPZ equation plays in the 
study of stochastic dynamic processes. The wide variety of dynamic 
processes governed by the KPZ equation form the so-called KPZ 
universality class---a class of seemingly unrelated dynamics whose 
bulk properties obey, on a course-grained level, this one master 
equation. One member of the KPZ 
universality class, the well studied one-dimensional asymmetric 
exclusion process (ASEP) describes a driven lattice gas with hard 
core exclusions~\cite{spohn91}, and has also been applied to studies 
of highway traffic~\cite{raje99,beli01}, protein synthesis~\cite{shaw03}, 
and sequence alignment~\cite{bund00,bund02,RECOMB05}. Characterizing 
the underlying properties of the ASEP promises a greater 
understanding for these specific studies as well as insights into 
the broader KPZ universality class. Not surprisingly, much effort has 
been spent calculating some of the many properties of the ASEP such 
as the density profile, steady states, mass gaps, and diffusion 
constants~\cite{derr98a,derr04,krug91,meak86,kim95,dhar87,gwa92,derr93,
stin95,jano92}. Though many questions still remain, these myriad studies 
have helped to uncover a great deal of insight into the ASEP.

In their study of the ASEP, Derrida and Lebowitz~\cite{derr98} extended 
the Bethe Ansatz approach of Gwa and Spohn~\cite{gwa92} in order to solve 
the totally ASEP for particle displacement in the asymptotic limit of 
large system size~\cite{derr98}. 
One of the most interesting aspects of their solution involves the 
scaling function $\DLSF$, describing the non-linear behavior of the 
total particle flux. The Derrida-Lebowitz scaling function (DLSF) is 
independent of any model parameters. 
Thus, this scaling function was conjectured to be {\it universal}, 
i.e., characteristic of all KPZ systems. In a follow-up study, 
Derrida and Appert analytically continued the Derrida-Lebowitz 
scaling function (DLSF), successfully completing the solution for all 
space in the asymptotic limit of large system sizes~\cite{derr99}.
Since then, a large number of studies have given strong
evidence for the universality
of the DLSF. These studies fall into two classes. On the one hand,
for a few closely related variants of the continuous time totally 
ASEP~\cite{lee99,brun00,povo03} the
characteristic DLSF behavior has been analytically verified.
On the other hand, numerical studies have bolstered the universality claim
of the DLSF~\cite{derr99,lee99,povo03,appe00,weic00} for a much broader
range of systems including genuinely {\em discrete} time systems. However,
these numerical methods do not directly verify the universality of
the DLSF, but rather verify the universality of certain cumulant ratios
which must be universal if the DLSF is universal~\cite{derr99}. These numerical
approaches cannot directly verify the DLSF since they use {\em sampling}
methods that are inherently unable to probe the full DLSF which 
contains information about {\em statistically rare} events. In addition
to that, both analytical and numerical studies rely on examining the 
behavior of the so-called scaling region~\cite{derr99} which applies
only in the limit of very large system sizes ($\sim 1280$), 
especially in the discrete time case ($> 10240$).

By examining properties applicable to the intermediate scaling region, 
we created a method for {\em directly} measuring the DLSF for
discrete time systems of considerably smaller size ($< 22$). Our method
works without resorting to the stochastic sampling that makes cumulant
methods so time consuming. As an application, we show that the discrete 
time ASEP with both single and multiple particles per site under any 
parameter choice is characterized by the same DLSF conjectured to be 
universal for all processes within the KPZ universality class. Our
method can also be used to study the particle hopping in other ASEP 
scenarios, including the open boundary ASEP. Once DLSF behavior has been 
verified for any given system, our method also provides a way to 
numerically calculate the non-universal scaling constants that, together 
with the known form of the DLSF, can be used to calculate properties such 
as the particle hopping distribution or the large deviation function.

This paper will begin with a short review of the continuous time single
particle system from which Derrida {\it et al.}~\cite{derr98,derr99}
initially derived the DLSF in section~\ref{sec:review}. Then,
we will reveal a small extension of the Derrida {\it et al.} results
that will allow us to measure the DLSF for the discrete time ASEP.
Next, we describe our method and its use with the discrete-time 
single- and multi-particle ASEP 
as an example of the application of our method. Last, we will discuss
the applicability of this method to other systems of interest.

\section{Review of the Asymmetric Exclusion Process}\label{sec:review}

\begin{figure}
\begin{center}
\includegraphics{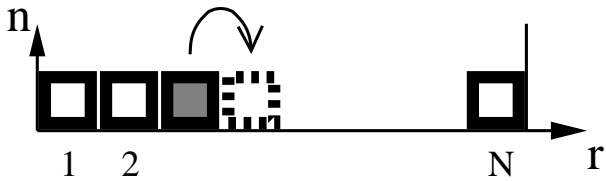}
\end{center}
\caption{\label{singleparticle}Diagram of a periodically bound totally 
ASEP. Only the particle shaded in gray, on site 3, may move to the space
on the right, outlined by a dotted line, with a probability $d\time$.
The remaining particles are prevented from moving by an adjacent particle
occupying the space to the immediate right. More explicitly, the particle
on site 2 finds it's advance blocked by the particle that resides on site
3. Meanwhile, the particle on site 1 is likewise blocked by the particle
on site 2. The particle on site N also finds it's movement thwarted by 
the particle on site 1 due to the periodic boundary conditions.}
\end{figure}

Derrida {\it et al.}~\cite{derr98,derr99} examine a periodically bound 
continuous time single particle totally  ASEP with $\size$ sites each 
capable of holding one particle as shown in Fig.~\ref{singleparticle}. 
These particles may move only to the right with a probability $d\time$ 
if the target site is unoccupied. The total number of particles in the 
system is fixed at $\totpartnum$.

In order to characterize the probability distribution $P(\position)$
of the cumulative particle hopping per site $\position$, they
calculate the generating function $\GF$, given by
\begin{equation}
\GF = \langle\exp[\gam \position]\rangle
\end{equation}
where the brackets $\langle\cdot\rangle$ denote the ensemble average.
In order to solve for this generating function, they use
the large time $\time$ behavior~\cite{derr98},
\begin{equation}
\langle\exp[\gam \position]\rangle \sim \exp[\lambdamax(\gam)\time]
\end{equation}
where $\lambdamax(\gam)$ is the largest eigenvalue of the modified
transfer matrix $\bigT$. If so inclined, the reader may refer to 
Appendix~\ref{sec:lambdamax} to examine the full solution to 
$\lambdamax(\gam)$, originally obtained by Derrida and Lebowitz 
in~\cite{derr98} and analytically continued by Derrida and Appert 
in~\cite{derr99}. The most important feature of this solution for
$\lambdamax(\gam)$ is that it has the scaling behavior
\begin{equation}
\lambdamax(\gam) - \gam\lconstD 
= \frac{\aD \DLSF(\gam\size^{1/2}\bD)}{\size^{3/2}}
\label{eq:linear}
\end{equation}
for fixed filling ratio $\totpartnum/\size = \partdensity$ in the scaling 
limit $\size\to\infty$ with $\gam\size^{1/2}$ held constant. $\aD,\bD$ 
and $\lconstD$ are all constants that depend solely on the particle 
density $\partdensity$. The most interesting 
aspect of this solution is the Derrida-Lebowitz scaling function $\DLSF$, 
whose form is independent of all system parameters. This led~\cite{derr99} 
to postulate that, in fact, $\DLSF$ represents a {\em universal scaling 
function} that can be used to describe {\em all} systems within the KPZ
universality class. For details about the explicit representation of
the Derrida-Lebowitz scaling function, the reader may refer
to Appendix~\ref{sec:DLSF}.

\section{Scaling Behavior for Intermediate N}

\begin{figure}
\begin{center}
\includegraphics{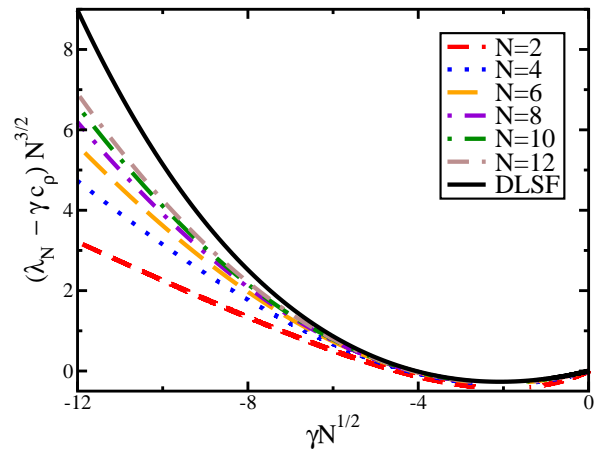}
\end{center}
\caption{\label{OLDda}Plot of the left hand side of 
Eq.~(\ref{eq:linear}) rescaled by $\size^{3/2}$ versus $\gam \size^{1/2}$ for
variously sized systems. 
A plot of the appropriately rescaled DLSF $\DLSF$ is included for reference.
The curves for smaller $\size$ do not agree well with the DLSF. This tells
us that the left hand side of Eq.~(\ref{eq:linear}) cannot be described by
$\DLSF$ alone for small system sizes. The gap 
in the curves is a consequence of the dramatic increase in computational
effort in its vicinity due to the slow convergence of the infinite series
defining the solutions $\lambdamax(\gam)$ close to their radii of
convergence (see appendix~\ref{sec:lambdamax}).}
\end{figure}

Eq.~(\ref{eq:linear}) contains a formula for successfully describing
$\lambdamax(\gam)$ in the scaling regime of constant $\gam\size^{1/2}$
in the limit of large system sizes. However, when considering small
systems or large $\gam$ at a fixed system size, Eq.~(\ref{eq:linear})
no longer adequately describes $\lambdamax(\gam)$. Fig.~\ref{OLDda}
shows just how poorly Eq.~(\ref{eq:linear}) performs under these
conditions. If one wanted to extract the scaling function $\DLSF$ from
numerical data for $\lambdamax(\gam)$ in order to, e.g., verify its
universality in a different system, one would have to go to very large
system sizes $\size$.

\begin{figure}
\begin{center}
\includegraphics{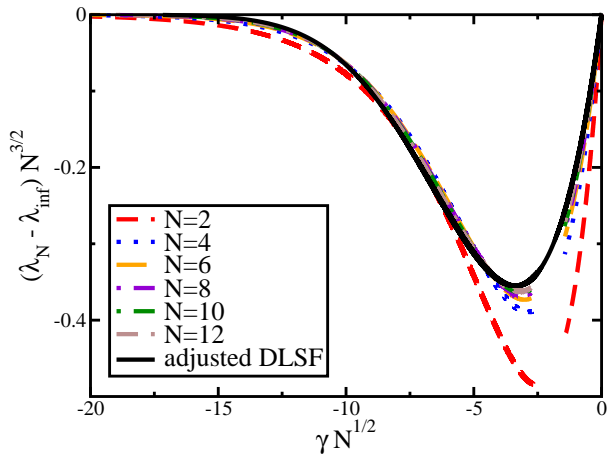}
\end{center}
\caption{\label{da}Plot of the left hand side of Eq.~(\ref{eq:newform}), 
rescaled by $\size^{3/2}$, for systems size $N = 2,4,6,8,10$ and $12$. Notice 
how, even in the case of very small sizes, these values plotted are 
almost indistinguishable from not only each other, but the adjusted
DLSF, given in Eq.~(\ref{eq:adlsf}) which they 
are plotted against. The gap in the data occurs due to the slow 
convergence of the plotted solutions (discussed in 
appendix~\ref{sec:lambdamax}) 
around the point $\B_c$ (as given by Eq.~(\ref{eq:Bc}))}
\end{figure}

Here, we introduce another approach that can be utilized to directly
verify the scaling behavior of $\lambdamax(\gam)$. To this end, we
exploit the fact that $\lambdamax(\gam)$ not only has a well-defined
scaling limit as $\size\to\infty$ at constant $\gam\size^{1/2}$ (which
implies $\gam\to0$) but that there is also a well-defined scaling limit
as $\size\to\infty$ at {\em constant} $\gam$ for $\gam<0$. This
scaling limit is given~\cite{derr99} by
\begin{equation}
\lim_{\size\to\infty}\lambdamax(\gam)\equiv \lambdaneg(\gam) =
\frac{-(1-\exp[\gam\partdensity])(1-\exp[\gam(1-\partdensity)])}
{1-\exp[\gam]}\label{eq:lambdaneg}.
\end{equation}
This behavior is {\em not} universal, as the explicit dependence
of the result on the particle density $\partdensity$ shows. However,
the existence of such a scaling limit should still be a universal feature
of all KPZ systems.

We propose that a {\em combination} of the two scaling
limits in Eqs.~(\ref{eq:linear}) and~(\ref{eq:lambdaneg}) provides a
faithful representation of the full function $\lambdamax(\gam)$ for $\gam<0$
and very small system sizes $\size$. More precisely, we suggest that
\begin{equation}
\lambdamax(\gam) - \lambdaneg(\gam) 
\approx \frac{\aD \DLSF(\gam\size^{1/2}\bD)}{\size^{3/2}} 
+ \frac{\aD (\gam \bD)^3}{24\pi}\label{eq:newform}
\end{equation}
already for very moderately sized $\size=O(10)$. This agreement is
essentially due to the fact that $\lambdaneg(\gam)$ provides an
approximation the $\lambdamax(\gam)$ behavior that is far superior to
the linear approximation from which the Derrida-Lebowitz scaling
function $\DLSF$ was originally derived.

It is instructive to verify that our new relation
Eq.~(\ref{eq:newform}) indeed simplifies to the two known scaling
relations Eqs.~(\ref{eq:linear}) and~(\ref{eq:lambdaneg}) in the
appropriate limits. For fixed $\gam<0$ the argument of the
Derrida-Lebowitz scaling function $\DLSF$ on the right hand side goes
to negative infinity as $\size\to\infty$. In this limit it is
known~\cite{derr99} that
\begin{equation}
\DLSF(\beta) \simeq -\frac{\beta^3}{24\pi} =
-\frac{(\gam \size^{1/2} \b)^3}{24\pi}\label{eq:Gneg}
\end{equation}
where the sub-leading terms decay exponentially with $|\beta|$. Since
this leading term in $\DLSF$ cancels with the second term on the right
hand side of Eq.~(\ref{eq:newform}), the right hand side vanishes as
$\size\to\infty$ and Eq.~(\ref{eq:lambdaneg}) emerges. If on the other
hand $\gam\size^{1/2}$ is held constant as $\size\to\infty$, $\gam$
vanishes. Thus, the function $\lambdaneg(\gam)$ can be expanded for
small $\gam$. Since $\lambdaneg(\gam)$ is an odd function, this
expansion contains only terms with odd powers of $\gam$. The linear
term precisely yields the term $\gam\lconstD$ in
Eq.~(\ref{eq:linear}). The $\gam^3$ term equals the second term on the
right hand side of Eq.~(\ref{eq:newform}) (note that this implies a
non-trivial connection between the non-universal constants $\aD$,
$\bD$, and $\lconstD$ and the non-universal function
$\lambdaneg(\gam)$). All higher order terms in $\gam$ become
sub-leading as $\size\to\infty$ in this scaling limit. Thus,
Eq.~(\ref{eq:linear}) reemerges. As it becomes clear from this
discussion, the additional term on the right hand side of
Eq.~(\ref{eq:form}) can either be understood as a $\gam^3$ correction
to the left hand side or as a correction to the DLSF $\DLSF$ on the
right hand side. It is the only term the scaling behavior of which
allows it to be interpreted as both a part of the universal scaling
function as well as a part of the infinite size solution $\lambdaneg(\gam)$.
For the purpose of this paper we will integrate this term into
our scaling function and call the function
\begin{equation}
\ADLSF(\beta)\equiv G(\beta) + \frac{\beta^3}{24\pi}\label{eq:adlsf}
\end{equation}
the adjusted Derrida-Lebowitz scaling function.

Note that while Eq.~(\ref{eq:newform}) has only been shown to hold
in the two limiting cases discussed above, in actuality, it works
exceedingly well even for small systems. Fig.~\ref{da} shows the left
hand side of Eq.~(\ref{eq:newform}), rescaled by $\size^{3/2}$, 
plotted for $\size = 2,4,6,8,10$ and $12$. The right hand side of
Eq.~(\ref{eq:newform}) is plotted as well, with the appropriate
rescaling factors given by~\cite{derr99}. The quality with which 
Eq.~(\ref{eq:newform}) captures the behavior of this intermediate
scaling region is suprising, especially when considering the small
system sizes. This empirical observation forms the foundation of the 
methodology presented herein for numerically 
understanding the behavior of quantities, such as the particle hopping, 
in systems that currently remain beyond the reach of purely analytic methods.

\section{Discrete Time Asymmetric Exclusion Process}

\begin{figure}
\begin{center}
\includegraphics{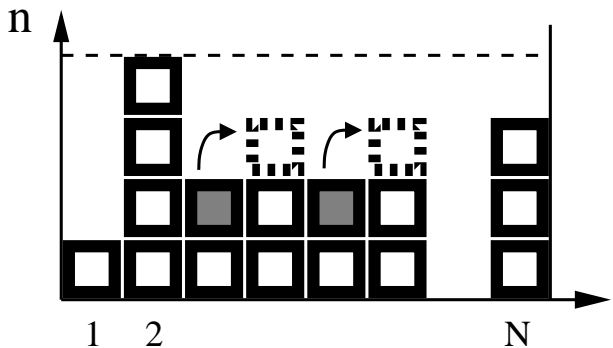}
\end{center}
\caption{\label{particles}Diagram of our discrete time asymmetric 
exclusion process. The diagram shows an odd time where only particles
on even sites are considered for hopping. Notice that the particle at
site 1 cannot move because site 2 is already at maximum occupancy.
However, for sites 3 and 5, the shaded particles can hop to the right
with a probability $\hop$, as shown by the boxes with dotted outlines.}
\end{figure}

In the following, as a specific application of our method, we verify the
universality of the DLSF for  
the discrete time totally ASEP using the sublattice-parallel updating 
scheme~\cite{raje99}. In our totally ASEP, we consider a periodically 
bound system of size $\size$ where $\totpartnum$ particles can only 
move to the right, 
and each site can accommodate up to $\sitepartnum$ particles.
During each odd time interval $\time$, the positions 
with odd numbers are evaluated for transitions. Particles 
can hop only if available space exists to the right for the particle 
to move, i.e., only if the site does not already contain the maximum 
number of allowed particles $\sitepartnum$. For allowed transitions, 
particles hop with probability $\hop$ and stay put with probability 
$1-\hop$. If there is more than one particle on a site only one 
particle is considered for hopping. For even time intervals, the exact 
same dynamic occurs at the even numbered positions. Notice that, in 
order to use this sublattice-parallel update scheme, the number of sites 
$\size$ must be even. For this system, we solve via a transfer matrix 
method described in Appendix~\ref{sec:matrix} for the exponential term, 
$\lambdamax$, of the generating function, $\GF$, of the total particle 
flux per site, $\position$, in the large time limit, $\time\to\infty$. 
Again, let us first define this generating function as
\begin{equation}
\GFall\equiv\langle
\exp[\gamma \position]\rangle
\label{eq:Z}
\end{equation}
where the brackets $\langle\cdot\rangle$ denote the ensemble average 
and $\parameters$ summarizes the specific parameters contributing to 
the evaluation of the particle flux per site $\position$, 
i.e., $\hop,\totpartnum$, and $\sitepartnum$. As in the continuous
time case examined by Derrida and Lebowitz \cite{derr98}, the 
generating function behaves like
\begin{equation}
\GFall\sim \exp[\fall\time]
\end{equation}
for large times, where $\fall$ is the largest eigenvalue of the 
characteristic matrix, 
$\bigT_\size(\gam;\parameters)$, the technical details of whose 
construction are provided in Appendix~\ref{sec:matrix}. 
%
%
%
%
%

Since we will only be able to numerically evaluate $\fall$ for
relatively small $\size$, we apply the method presented in the previous
section. 
Toward this end, we introduce the 
function $\Fall$ describing the infinite size behavior of the $\fall$
as
\begin{equation}\label{eq:defPhi_W}
\Fall=\lim_{\size\to\infty}\fall.
\end{equation}
The new scaling form then becomes
\begin{equation}\label{eq:form}
\fall - \Fall = 
\frac{\a \ADLSF(\gam \size^{1/2} \b)}{\size^{3/2}}.
\end{equation}

\begin{figure}
\begin{center}
\includegraphics{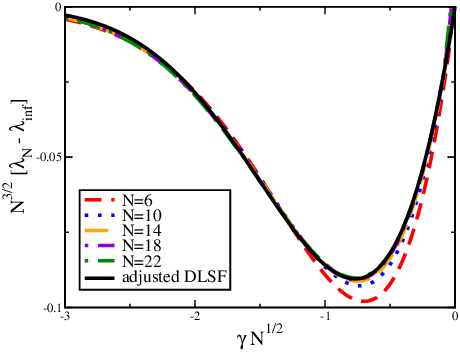}
\end{center}
\caption{\label{fig1}The appropriately rescaled and adjusted DLSF 
(in accordance with the right hand side of Eq.~(\ref{eq:form1})) 
plotted against the direct measurement of the left hand side of 
Eq.~(\ref{eq:form1}) for $\sitepartnum=1$, 
$\partdensity=\totpartnum/(\sitepartnum\size)=1/2$ and $\hop=3/4$. 
The left hand side of Eq.~(\ref{eq:form}) has been plotted for 
$\size=6,10,14,18,$ and $22$. It becomes difficult to distinguish 
the values for the larger widths from the adjusted DLSF solved
in~\cite{derr99} since they lie in near perfect agreement.
This indicates that the proposed universality of the DLSF does 
hold for this system. The scaling factors $\a$ and $\b$ in 
Eq.~(\ref{eq:form}), neither of which depends on the finite size 
effects, are chosen only once for the largest system, $\size=22$, 
and control the scaling of the adjusted DLSF.}
\end{figure}

For the specific case in which the allowed number of particles 
per site equals one, $\sitepartnum=1$, and the system is half filled,
$\totpartnum=\size/2$, there exists an exact analytical 
solution~\cite{bund00,bund02} for $\Fall$.
\begin{equation}\label{eq_lambdaresult}
\f_\infty(\gam;\parameters_{\sitepartnum=1,\totpartnum=\frac{\size}{2}})
=\ln\left(\frac{\sqrt{\hop}+\exp[-\gam]}
{1+\sqrt{\hop}\exp[-\gam]}\right)
\end{equation}
This, combined with our numerical transfer matrix method for 
calculating $\fall$, allows us to directly measure the left hand 
side of Eq.~(\ref{eq:form}). Since the adjusted DLSF on the right hand 
side has already been solved, only the scaling coefficients $\a$ and
$\b$ are unknown. This allows us to numerically calculate the left 
hand side of Eq.~(\ref{eq:form}) and use the results to fit $\a$ and 
$\b$ for one (the largest) system size $\size$. For convenience, 
here we multiply Eq.~(\ref{eq:form}) by $\size^{3/2}$
\begin{equation}\label{eq:form1}
\begin{array}{l}
\size^{3/2}(\fall - \Fall) =
\a \ADLSF(\gam \size^{1/2} \b).
\end{array}
\end{equation}
and plot the left hand side of Eq.~(\ref{eq:form1}) against 
$\beta\equiv\gam\size^{1/2}$ in Fig.~\ref{fig1} for 
$\size = 6,10,14,18,$ and $22$. The agreement between the various 
curves in Fig.~\ref{fig1} gives a good indication that the asymptotic 
solution for the DLSF indeed applies to our discrete time ASEP.

\section{Discrete Time Multi-Particle Asymmetric Exclusion Process}

\begin{figure}
\begin{center}
\includegraphics{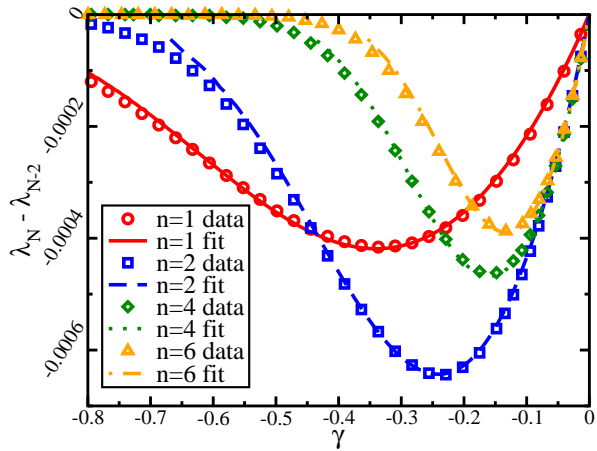}
\end{center}
\caption{\label{fig2}Plot of the right hand side of 
Eq.~(\ref{eq:rhorho}) versus the appropriately fitted left hand 
side for $\hop=3/4$ for various $\size$. This plots shows the
excellent agreement between the calculated data and the 
Derrida {\it et al.} proposed form of the scaling function
under different parameter choices. 
The total number of particles
has been chosen such that $\totpartnum=\size\sitepartnum/2$, i.e.,
a half-filled system. For $\sitepartnum=1$, $\size=22$ was used,
for $\sitepartnum=2$, $\size=14$, for $\sitepartnum=4$, $\size=10$,
and for $\sitepartnum=6$, $\size=8$.}
\end{figure}

In order to give a further example of the usability of our method, we
also apply it to the case where we allow more than one particle per site. 
A lattice point with more than one allowed state or highway traffic with 
more than one lane both provide very good examples of why such a multiple 
particle ASEP is important in and of itself. For $\sitepartnum>1$, an 
analytic solution for $\Fall$ does not exist. Thus, we cannot quite follow 
the method shown in the previous sections. However, 
we can still test both the validity of the DLSF, and calculate the
non-universal scaling factors through other means. Since we have a method 
for obtaining $\fall$, we take the difference
\begin{equation}
\begin{array}{l}
\fall - \f_{\size-2}(\gam;\parameters) = \\
\frac{\a \DLSF(\gam \size^{1/2} \b)}{\size^{3/2}} - 
\frac{\a \DLSF(\gam (\size-2)^{1/2} \b)}{(\size-2)^{3/2}}
\label{eq:rhorho}
\end{array}
\end{equation}
in order to eliminate the need for the value of $\Fall$. Notice that
the size independent terms ($\a(\gam\b)^3/(24\pi)$) cancel out, leaving us
with the original DLSF $\DLSF$. Once
again, we find the scaling coefficients $\a$ and $\b$ by fitting 
the right hand side to the left hand side of Eq.~(\ref{eq:rhorho})
for a single $\size$. 
Fig.~\ref{fig2} shows the right hand side of Eq.~(\ref{eq:rhorho})
for the largest $\size$
with the numerically fitted values for $\a$ and $\b$ plotted against
the left hand side whose value is obtained using the same
transfer matrix method described in appendix~\ref{sec:matrix}.
While in this approach curves obtained for different $\size$ do not
overlap each other, we note that we obtain similarly good agreement
as in Fig.~\ref{fig2} for small system sizes $\size$ {\em without}
refitting the scaling parameters $\a$ and $\b$ (data not shown).
This once again results in excellent agreement between our solution
for the left hand side and the properly rescaled right hand side 
for various parameter values $\parameters$ including different values 
for $\sitepartnum$, $\hop$, and $\size$. 

Fig.~\ref{fig1} and Fig.~\ref{fig2} each display excellent agreement 
with established KPZ theory and come without any real adjustment or 
reformulation of existing theory. They are simply the direct result 
of studying our multi-particle discrete time ASEP dynamics modeled 
by a transfer matrix method. Once $\a$ and $\b$ have been determined, 
their values can be utilized in order to numerically compute the 
infinite form $\Fall$ from $\fall$ for even small system sizes by 
utilizing Eq.~(\ref{eq:form})---a fact which is particularly important 
in cases where $\Fall$ remains unsolved. Of course, given a form
for $\Fall$, even though it be numerical, allows one to calculate
properties of the system in the thermodynamic limit, such as the 
particle hopping and large deviation function.
 
In summary, the fast convergence in $\size$ allows for rapid 
calculation of an entire function $\Fall$, an otherwise difficult 
quantity to compute for most systems. The accessibility of $\Fall$ 
through studies of smaller systems has already been put to use in 
calculating practical quantities important in sequence 
alignment~\cite{RECOMB05}.

\section{Conclusion}
In this letter, we have presented a numerical 
method that allows a direct numerical calculation of the universal DLSF 
and it's non-universal scaling factors. In contrast to previous numerical 
studies, our method does not rely on sampling but rather on the exact 
diagonalization of sparse modified transfer matrices for very small finite 
size systems. Thus, it is able to capture the full information on rare events 
that the DLSF encodes.
As an application, we have extended the universality of the
Derrida-Lebowitz scaling function to inherently discrete time hopping
processes with multiple particles per site by directly measuring the
scaling function itself. We found that the numerically determined 
scaling function converge toward the universal scaling function already 
for relatively small finite systems. 

The method we have outlined is applicable to a variety of discrete KPZ
systems, including the important open boundary ASEP and the open
boundary partially ASEP used for modeling true non-equilibrium driven
lattice gas dynamics. In addition to this, the more complex interactions 
of multiple number and types of particles can be modeled using our
method. 

\section{Acknowledgments}

RB gratefully acknowledges the funding from the National
Science Foundation through grants DBI-0317335 and DMR-0404615.

\begin{appendix}
\section{$\lambdamax$ for a Continuous Time Totally Asymmetric
Exclusion Process}\label{sec:lambdamax}

The solution for $\lambdamax(\gam)$ can be given by the parametric
equation~\cite{derr98}
\begin{eqnarray}
\lambdamax(\gam) &=& -\totpartnum\sum_{\q=1}^\infty \B^\q 
\frac{(\size\q-2)!}{(\totpartnum\q)! (\size\q - \totpartnum\q - 1)!}
\label{eq:lambda_zero}\\
\gam &=& -\size\sum_{\q=1}^\infty \B^\q 
\frac{(\size\q-1)!}{(\totpartnum\q)! (\size\q - \totpartnum\q)!}
\label{eq:gam_zero}
\end{eqnarray}
for $-\B_c < \B < \B_c$, where the radius of convergence $\B_c$ is 
given by
\begin{equation}
\B_c = \frac{\totpartnum^\totpartnum 
(\size - \totpartnum)^{(\size - \totpartnum)}}
{\size^\size}.\label{eq:Bc}
\end{equation}
This set of equations represent the solution for $\lambdamax(\gam)$
in the region $\gam_- < \gam < \gam_+$, where
\begin{eqnarray}
\gam_- &=& -\size\sum_{\q=1}^\infty (\B_c)^\q 
\frac{(\size\q-1)!}{(\totpartnum\q)! (\size\q - \totpartnum\q)!}
\label{eq:gamneg}\\
\gam_+ &=& -\size\sum_{\q=1}^\infty (-\B_c)^\q 
\frac{(\size\q-1)!}{(\totpartnum\q)! (\size\q - \totpartnum\q)!}.
\label{eq:gampos}
\end{eqnarray}
Derrida and Appert term this the 'scaling region', i.e., the region
where Eq.~(\ref{eq:linear}) holds true in the limit $\size\to\infty$.

In the strongly negative region, $\gam < \gam_-$, the solution 
is given by the analytic continuation of Eqs.~(\ref{eq:lambda_zero})
and~(\ref{eq:gam_zero}), found in~\cite{derr99},
\begin{equation}
\begin{array}{lll}
\lambdamax(\gam) &=& 
\dfrac{\frac{1}{2\B^{1/p}}(1-2\B^{1/p}-\sqrt{1-4\B^{1/p}})}
{1 + \frac{1}{2\B^{1/p}}(1-2\B^{1/p}-\sqrt{1-4\B^{1/p}})}\\[5mm]
&-& \dfrac{\frac{1}{2\B^{1/p}}(1-2\B^{1/p}+\sqrt{1-4\B^{1/p}})}
{1 + \frac{1}{2\B^{1/p}}(1-2\B^{1/p}+\sqrt{1-4\B^{1/p}})}\\[5mm]
&-&\totpartnum\sum_{\q=1}^\infty \B^\q 
\dfrac{(\size\q-2)!}{(\totpartnum\q)! (\size\q - \totpartnum\q - 1)!}
\label{eq:lambda_neg}
\end{array}
\end{equation}
\begin{equation}
\begin{array}{lll}
\gam &=& \frac{\size}{\totpartnum}\{
\ln({1 + \frac{1}{2\B^{1/p}}(1-2\B^{1/p}-\sqrt{1-4\B^{1/p}})})\\[5mm]
&-& \ln({1 + \frac{1}{2\B^{1/p}}(1-2\B^{1/p}+\sqrt{1-4\B^{1/p}})})\}\\[5mm]
&-& \size\sum_{\q=1}^\infty \B^\q 
\dfrac{(\size\q-1)!}{(\totpartnum\q)! (\size\q - \totpartnum\q)!}
\label{eq:gam_neg}
\end{array}
\end{equation}
for $0 < \B < \B_c$.

\section{The Derrida-Lebowitz Scaling Function}\label{sec:DLSF}

The form of the Derrida-Lebowitz scaling function can be given as
\begin{eqnarray}
\DLSF(\beta) &=& \frac{4}{3 \sqrt{\pi}} \int_{0}^\infty \epsilon^{3/2}
\frac{C e^{-\epsilon} d\epsilon}{1 + C e^{-\epsilon}}
\label{eq:G_pos}\\
\beta &=& \frac{2}{\sqrt{\pi}} \int_{0}^\infty \epsilon^{1/2}
\frac{C e^{-\epsilon} d\epsilon}{1 + C e^{-\epsilon}}
\label{eq:beta_pos}
\end{eqnarray}
for $C > -1$. For the region less than 
$\beta_- \equiv \lim_{C\to -1}\beta$, the analytic continuation of 
$\DLSF$ is given by
\begin{eqnarray}
\DLSF(\beta) &=& \frac{8}{3} \sqrt{\pi} \left[ -\ln(-C) \right]^{3/2} -
\sum_{q=1}^\infty (-C)^q q^{-5/2}
\label{eq:G_neg}\\
\beta &=& -4\sqrt{\pi} \left[ -\ln(-C) \right]^{1/2} - 
\sum_{q=1}^\infty (-C)^q q^{-3/2}
\label{eq:beta_neg}
\end{eqnarray}
for $0<C<1$.

\section{Brief Description of Matrix Method}\label{sec:matrix}
We shortly describe how we actually 
calculate $\fall$. In our brand of the ASEP, we use a sublattice-parallel 
updating scheme where hopping probabilities in even and odd time intervals 
are evaluated separately. Because our discrete time processes may be
thought of as a combination of a number of very simple process occurring
in some sequential order, we may first examine the base dynamic
and expand this into the larger picture. Before examining the 
movement at all positions, we study the dynamics of a single 
hopping transition. We first create the transfer matrix 
$\smallT(\gam=0;\parameters)$ describing the transition probabilities for 
one pair of sites~\cite{bund02} using particle occupancy number
$\d$
as our basis. Next, we modify this matrix by multiplying all off 
diagonal elements by the factor $\exp [ - \gam / \size]$. This 
effectively tags the average number of hops per site. For 
$\sitepartnum=2$ particles per site this results in the matrix
\begin{equation}
\smallT(\gam;\parameters_{\sitepartnum=2})=\left(
\begin{array}{ccccccccc}
1 & 0 & 0 & 0 & 0 & 0 & 0 & 0 & 0\\[1mm]
0 & \hop & 0 & 0 & 0 & 0 & 0 & 0 & 0\\[1mm]
0 & 0 & \hop & 0 & 0 & 0 & 0 & 0 & 0\\[1mm]
0 & \c & 0 & 1 & 0 & 0 & 0 & 0 & 0\\[1mm]
0 & 0 & \c & 0 & \hop & 0 & 0 & 0 & 0\\[1mm]
0 & 0 & 0 & 0 & 0 & \hop & 0 & 0 & 0\\[1mm]
0 & 0 & 0 & 0 & \c & 0 & 1 & 0 & 0\\[1mm]
0 & 0 & 0 & 0 & 0 & \c & 0 & 1 & 0\\[1mm]
0 & 0 & 0 & 0 & 0 & 0 & 0 & 0 & 1\\[1mm]
\end{array}  \right)
\end{equation}
in the basis (00, 01, 02, 10, 11, 12, 20, 21, 22)
where $\c=(1-\hop)\exp[-\gam/\size]$. 

From this smaller transfer matrix, we may build up this 
single pair description into the larger $\size$-site 
picture by taking the tensor product
\begin{equation}
\bigT_\size(\gam;\parameters) = \bigotimes_{k=1}^{\size/2} 
\smallT(\gam;\parameters)
\end{equation}
and eliminating any states that do not contain the right number of
particles \totpartnum.
This gives us the matrix that models particle hopping from sites 
in one time interval. Converting this to the appropriate basis for 
the next time interval can be done by utilizing the translation 
operator $\C$ defined such that
\begin{equation}
\C|\d_0 \d_1 \ldots \d_{\size-1}\rangle
\equiv|\d_1 \d_2 \ldots \d_{\size-1} \d_0\rangle.
\end{equation}
After exploiting translational invariance and up-down mirror symmetry
in order to reduce the size of our state space we obtain the additional 
identity $\C = \C^{-1}$ on this reduced state space.
Then the matrix product $\bigT_{even} \bigT_{odd} = 
\bigT_\size (\C^{-1} \bigT_\size \C)=(\bigT_\size \C)^2$ 
describes the particle hopping of our discrete time ASEP for all
$\size$ sites. These dynamics can be viewed as a Markov 
process on a $(\sitepartnum+1)^{\size}$--dimensional state space of the 
equal time difference vector 
$|\d(0,\time),\d(1,\time),\ldots,\d(\size,\time)\rangle$. 
Solving for the largest eigenvalue $\eigen_\size(\gam;\parameters)$ 
of $\bigT_N\C$ gives us the particle hopping function for finite 
size~\cite{bund00,bund02}
\begin{equation}\label{eq:eigen}
\fall = \ln \eigen_\size(\gam;\parameters).
\end{equation}
This is the function used in order to produce the results given above.
Calculating this largest eigenvalue is somewhat challenging, since
this matrix description grows very quickly with $\size$. However,
the matrices are very sparse with the number of non-zero matrix 
elements growing almost linearly with the matrix dimension. This
makes it possible to numerically obtain the largest eigenvalue
required in Eq.~(\ref{eq:eigen}) using the implicitly restarted
Arnoldi method~\cite{ARPACK} for matrix dimensions up to around 
$10^5$. This allows us to plot the entire function $\fall$ in very
reasonable timescales.

\end{appendix}

\end{document}